\documentclass[10pt,a4paper,twocolumn,english,floatfix,groupedaddress,
superscriptaddress,showpacs]{revtex4}

\usepackage{graphicx}
\usepackage{epsfig}
\usepackage[english]{babel}
\usepackage{amsmath}
\usepackage{amssymb}
\usepackage{amsfonts}
\usepackage{longtable}
\usepackage{dcolumn}
\usepackage{bm}
\usepackage{bbm}
\usepackage{color}

\begin{document}
\title{Vertex corrections in the dynamic structure factor in spin ladders}

\author{I. Exius}
\affiliation{Lehrstuhl f\"{u}r Theoretische Physik I, 
Technische Universit\"{a}t Dortmund,
 Otto-Hahn Stra\ss{}e 4, 44221 Dortmund, Germany}

\author{K. P. Schmidt}
\affiliation{Lehrstuhl f\"{u}r Theoretische Physik I, 
Technische Universit\"{a}t Dortmund,
 Otto-Hahn Stra\ss{}e 4, 44221 Dortmund, Germany}

\author{B. Lake}
\affiliation{Helmholtz Zentrum Berlin, Glienicker Stra\ss e 100, 14109 Berlin, 
Germany}
\affiliation{Institut f\"{u}r Festk\"{o}rperphysik, 
Technische Universit\"{a}t Berlin, Hardenbergstr.\ 36, 10623 Berlin, Germany}
\date{\textrm{\today}}

\author{D. A. Tennant}
\affiliation{Helmholtz Zentrum Berlin, Glienicker Stra\ss e 100, 
14109 Berlin, Germany}
\affiliation{Institut f\"{u}r Festk\"{o}rperphysik, 
Technische Universit\"{a}t Berlin, Hardenbergstr.\ 36,
10623 Berlin, Germany}
\date{\textrm{\today}}

\author{G. S. Uhrig}
\email{goetz.uhrig@tu-dortmund.de}
\affiliation{Lehrstuhl f\"{u}r Theoretische Physik I,
Technische Universit\"{a}t Dortmund,
 Otto-Hahn Stra\ss{}e 4, 44221 Dortmund, Germany}

\begin{abstract} 
We combine the results of perturbative continuous unitary transformations 
with a mean-field calculation to determine the evolution of the 
single-mode, i.e., one-triplon, contribution to the dynamic structure factor 
of a two-leg $S=1/2$ ladder on increasing temperature from zero to a finite
value. The temperature dependence is induced by two  effects:
(i) no triplon can be excited on a rung where a thermally activated
triplon is present; (ii) conditional excitation processes take place
if a thermally activated triplon is present.
Both effects diminish the one-triplon spectral weight upon heating.
It is shown that the second effect is the dominant vertex correction 
in the calculation of the dynamic structure factor.
The matrix elements  describing the conditional triplon excitation 
in the two-leg  Heisenberg ladder with additional four-spin ring exchange 
are calculated perturbatively up to  order 9.  The 
calculated results are compared to those of an inelastic neutron scattering 
experiment on the cuprate-ladder compound 
La$_{4}$Sr$_{10}$Cu$_{24}$O$_{41}$ showing 
convincing agreement for established values of the exchange constants.
\end{abstract}



\maketitle

\section{Introduction}\label{Intro}

Currently, the properties of strongly correlated systems
such as quantum spin systems are much better understood 
in their respective ground states, i.e., at zero temperature, 
than at finite temperatures. Generically, the ground state
is a pure state and the excitations may or may not have unconventional
statistics. But since they occur only in small numbers at zero
temperature their statistics is not exceedingly difficult to describe.

At finite temperatures, however, the physics becomes richer and more
complicated because there is a finite concentration of excitations.
Then their statistics and their mutual interactions play an important role.
Sharp $\delta$-peaks signalling infinitely long-lived excitations
acquire a broadening induced by thermal fluctuations. Spectral
weight is shifted due to variations of  the matrix elements and 
due to variations of the available phase space.

For one-dimensional quantum spin systems Essler and collaborators
analytically investigated the changes of line shapes in the dynamic 
structure factor \cite{james08,essle08,james09}. 
They considered systems where the elementary 
excitations are hardcore bosons which are local, i.e., 
they have minimal spatial extension. The main finding is that the line shapes 
broaden upon heating
as expected. The positions of the peaks shift depending upon the available
phase space. Rather unexpectedly, the line shapes at finite temperatures
are generically asymmetric, in agreement with earlier numerical evidence
\cite{mikes06}. Recently, the experimental verification of these
findings has been achieved \cite{tenna10a}.
So the usual expectation of Lorentzian broadening does not hold.
This is due to the interplay of the harcore property and the available
phase space.

In the present article, we aim at the investigation of the complementary
effect induced by vertex corrections. We focus on the changes in the
spectral weight of the elementary modes due to heating. We do not
study the broadening of the peaks corresponding to the elementary modes.

\begin{figure}[ht]
  \includegraphics[scale=0.25]{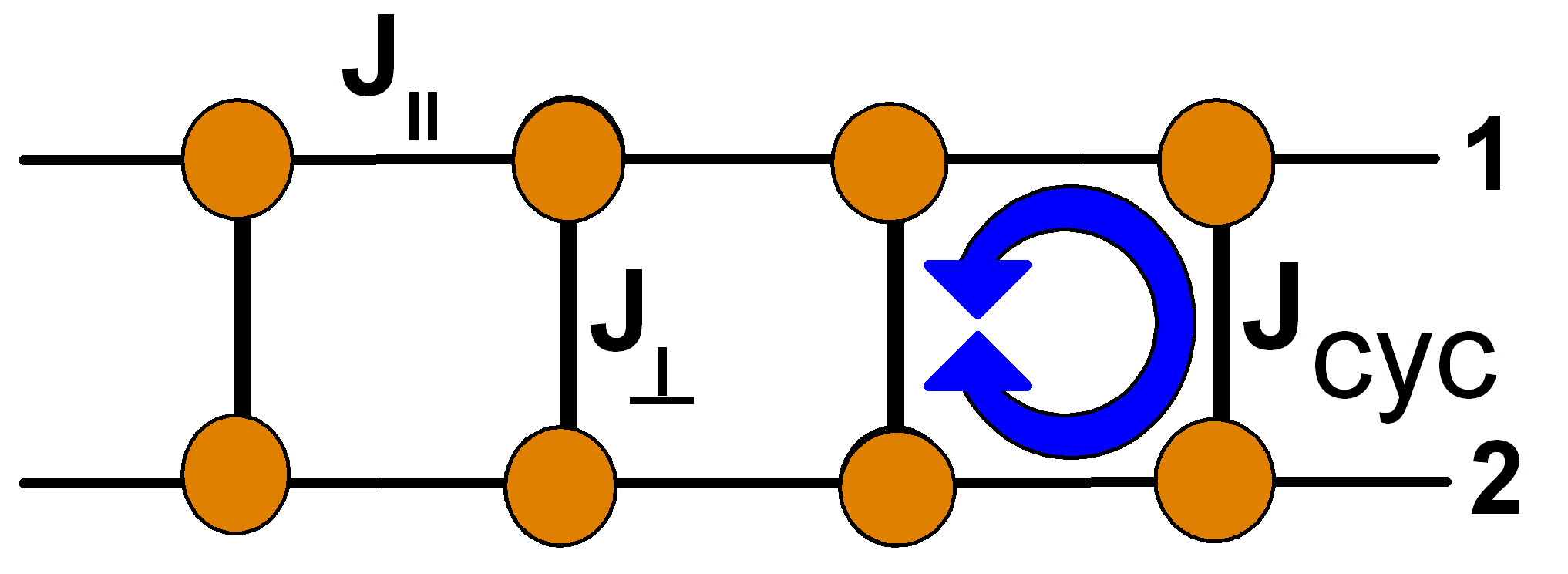}
  \caption{(Color online)
    Sketch of the two-leg spin ladder. Circles indicate spins with $S=1/2$. 
    Solid lines represent the two-spin exchange couplings on the rungs 
    $J_\perp$ and on the legs $J_\parallel$. 
    The four-spin cyclic exchange is denoted by $J_\textrm{cyc}$. }
  \label{leiterJ}
\end{figure}

The system we are investigating is a two-leg ladder
with localized $S=1/2$ at each vertex, see 
Ref.\ \onlinecite{schmi05b} and references therein. The dominant couplings are
the antiferromagnetic Heisenberg exchange couplings between nearest neighbours,
see Fig.\ \ref{leiterJ}. The nearest neighbour exchange couplings
are of similar magnitude. The ground state has zero spin and the elementary
excitations are of triplet character, i.e., they have $S=1$.
Thus we dub them triplons \cite{schmi03c} in order to distinguish them from the
elementary spin wave excitations, magnons,  of systems with magnetic long-range
order.  The centers of the triplons can be seen to sit on the rungs of the 
ladder. Further details about the system and its experimental realization
will be given in the following sections.

The temperature development of the spectral weight is induced by
two effects. First, the triplons are hardcore bosons so that
they can only be created on a rung if there is no other triplon
on this rung before. Clearly, this effect depends strongly on temperature
because the relative frequency of a successful excitation depends on
the concentration of thermally activated triplons. The higher the
temperature is the less probable is an excitation process. This effect,
due to the hardcore nature, has been dealt with in the
preceding investigations as well \cite{james08,essle08,james09}. 

Second, triplons can be excited \emph{if} there are other triplons
on rungs nearby. We call this process a conditional excitation
henceforth. Thereby, the matrix element for the excitation
of a triplon depends on the concentration of thermally activated
triplons. It turns out that the sign of the conditional excitations
is negative so that \emph{higher} temperatures and concomitantly higher
concentrations of triplons imply a \emph{lower} spectral weight.

The conditional excitations become more and more
important the more the triplons are extended. Hence the 
vertex corrections due to conditional excitations are significant
for isotropic spin ladders where the  couplings along rungs and legs
of the ladder are of similar magnitude. In the limits investigated 
previously the vertex corrections did not play a significant role
\cite{mikes06,james08,essle08}.

We find that the conditional excitations dominate the temperature 
dependence of the spectral weight. Our theoretical results agree
very  well with our experimental ones so that evidence for the
relevance of these vertex corrections is provided.

Technically, our theoretical calculations are based first on
continuous unitary transformations \cite{wegne94,kehre06} 
which map the original
model onto an effective model which conserves the number of
excitations \cite{uhrig98c,knett00a,knett03a}. This is done
in a perturbative fashion where the unperturbed Hamiltonian
is given by isolated rungs \cite{knett01b,knett03b}.
Alternative realizations of degenerate perturbation theory
are also based on systematic changes of the underlying basis
\cite{trebs00,zheng01a,oitma06}.

Second, we simplify the operators which describe
excitation processes by a mean-field (MF) treatment. 
In this step we focus on a description in terms of
single modes.

The paper is set up as follows. In Sect.\ \ref{M}
we present our model in detail. Subsequently, the theoretical approach is 
presented in  Sect.\ \ref{A}. The
theoretical results are shown in Sect.\ \ref{TR} and 
compared to experimental results in Sect.\ \ref{CE}.
Finally, the paper is concluded in Sect.\ \ref{Conclusions}.

\section{Model}
\label{M}

We study a two-leg $S= 1/2$ ladder as it is depicted in Fig.\ \ref{leiterJ}. 
The model is characterized by  Heisenberg exchange couplings on the rungs 
$J_\perp$ and on the legs $J_\parallel$ of the ladder as well as by
 a four-spin cyclic exchange $J_\textrm{cyc}$. The 
corresponding Hamiltonian in units of $J_\perp$ reads 
\begin{subequations}
\begin{eqnarray}
\frac{H}{J_\perp} &=&H_\perp+\frac{J_\parallel}{J_\perp}H_\parallel+
\frac{J_{\rm cyc}}{J_\perp} H_{\rm cyc}
\\
  &=& \sum_i \vec{S}_{i,l} \cdot \vec{S}_{i,r} + x \sum_{i; \tau}  
\vec{S}_{i,\tau} \cdot  \vec{S}_{i+1, \tau} + x_{\rm cyc} H_{\rm cyc},
\qquad
\label{hamil}
\end{eqnarray}
\end{subequations}
where $i$ denotes the rungs and $\tau \in \left\lbrace  1,2 \right\rbrace $ 
the upper and lower leg of the ladder. 
The expansion parameters employed are 
\begin{subequations}
\begin{eqnarray}
x &:=& J_\parallel/J_\perp
\\
x_\text{cyc} &:=& J_\text{cyc}/J_\perp.
\end{eqnarray}
\end{subequations}
The four-spin ring exchange is given by 
 \begin{eqnarray}
H_\textrm{cyc}&=& \sum_{i}  \left[\left( \vec{S}_{i, l}\cdot \vec{S}_{i+1,l}
\right) \left( \vec{S}_{i, r}\cdot \vec{S}_{i+1,r}\right)\right.
\nonumber\\
&&+ \left( \vec{S}_{i, l}\cdot \vec{S}_{i,r}\right)\left( \vec{S}_{i+1, l}
\cdot \vec{S}_{i+1,r}\right) 
\nonumber\\
&&\left.-\left( \vec{S}_{i, l}\cdot \vec{S}_{i+1,r}\right)
\left( \vec{S}_{i+1,l}\cdot \vec{S}_{i,r}\right)
\right].
\end{eqnarray}
These four-spin terms  constitute the most important correction 
to the commonly considered nearest neighbor 
Heisenberg exchange. By now, this is established for planes 
\cite{schmi90b,honda93,mulle02a,reisc04} and ladders
\cite{brehm99,mizun99,matsu00a,matsu00b,schmi05b}. The size of $J_{\rm cyc}$ 
in cuprate ladders was found to be of the order 0.2-0.25 $J_\perp$ 
\cite{nunne02,schmi03b,schmi05a,notbo07}. 
Further corrections like the two-spin 
interaction over the diagonal are almost one order of magnitude smaller 
\cite{mizun99} so that they are neglected in the following.

The physics of spin ladders is well understood from the limit of
isolated rungs $\{ x=0, x_{\rm cyc}=0\}$. 
In this limit the ground state is the product state 
of singlets and the excitations are completely local triplets. 
The elementary excitations 
at finite values $x$ and $x_{\rm cyc}$ are dressed triplets which are
called triplons \cite{schmi03c}. Triplon excitations are gapped and have
total spin one. In the following we will consider triplets on the
rungs as local triplons.

It is natural to represent the rung states in terms of bond 
operators \cite{chubu89a,chubu89b,sachd90}. 
Within this representation, the four-dimensional
 Hilbert space of each rung $i$ of the ladder is given by the singlet
\begin{equation}
 s^\dagger_{i}\vert 0 \rangle_i\!=\vert s \rangle_{i} \!=\frac{1}{\sqrt{2}}
\left( \vert\! \uparrow\downarrow\rangle -\vert\! \downarrow\uparrow\rangle
\right) 
\end{equation}
 and the three triplet states
\begin{subequations}
\begin{eqnarray}
t^\dagger_{-1,i}\vert 0 \rangle_i&\!=\vert t_{-1}  \rangle_{i} 
&\!=\frac{-1}{\sqrt{2}}\left( \vert\! \uparrow\uparrow\rangle -
\vert\!\downarrow\downarrow\rangle\right)
\\
t^\dagger_{0,i}\vert 0 \rangle_i&\!=\vert t_0 \rangle_{i} &\!=
\frac{\mathrm{i}}{\sqrt{2}}\left( \vert\! \uparrow \uparrow \rangle +
\vert\! \downarrow\downarrow\rangle\right)
\\
t^\dagger_{+1,i}\vert 0 \rangle_i&\!=\vert t_{+1} \rangle_i 
&\!=\frac{1}{\sqrt{2}}\left( \vert\! \uparrow\downarrow\rangle +
\vert\! \downarrow\uparrow\rangle\right).
\end{eqnarray}
\end{subequations}
The state $\vert 0 \rangle = \prod_i \vert 0 \rangle_i$ represents the
completely empty system. This is actually an unphysical state which
is needed here only for formal reasons.

The physically most important state is the ground state at 
$\{x=0,x_{\rm cyc}=0\}$
 which will be used as a reference state $|{\rm ref}\rangle$ for our
calculation. It is given by the product of singlets on all rungs
\begin{equation}
 \vert {\rm ref} \rangle:=\underset{i}{\prod}\vert s\rangle_i.
\end{equation}
One can also interpret $\vert {\rm ref}\rangle$ as a singlet 
condensate, i.e. singlet operators can be replaced by unity $s=1=s^\dagger$
Physically, $\vert {\rm ref} \rangle$
is the triplon vacuum, i.e., the state without any excitation.

The operator $t^\dagger_{\alpha,i}$ with 
$\alpha \in \left\lbrace0, \pm 1 \right\rbrace $ creates an elementary 
excitation with $S^z=\alpha$  (flavor) when applied to a singlet state on rung 
$i$. Each rung is either occupied by a singlet or by a triplon, resulting in 
the so called \emph{hardcore} constraint on all rungs $i$ 
\begin{equation}
 \mathbbm{1}=s_i^\dagger s^{\phantom{\dagger}}_i +\sum_\alpha 
t^\dagger_{\alpha,i} t^{\phantom{\dagger}}_{\alpha,i} \quad \alpha \in 
\left\lbrace -1,0,1\right\rbrace,
\label{bedingung}
\end{equation}
which implies particle-number conservation
in terms of singlets \emph{and} triplons. 

The Hamiltonian (\ref{hamil}) can be reformulated in terms of
the above bond operators.  One finds
\begin{equation}
 H = H_\perp + T_0 + T_{-2} + T_2 
\end{equation}
where the index $n$ in $T_n$ indicates the change of the total triplon number, 
e.g., $T_{+2}\propto t^\dagger t^\dagger\! ss$ are pair creation processes and 
$T_{-2}\propto tts^\dagger\! s^\dagger$ are pair annihilation processes. In 
contrast, $T_0$ does not change the number of triplons in the system. This 
term includes triplon hopping processes $t^\dagger ts^\dagger\! s$ and 
two-triplon interactions $t^\dagger t^\dagger t t$. Note that we have omitted 
the spatial and flavor index here in order to keep the notation light. 
Finally, the operator $H_\perp$ counts the total number of triplons in 
the system.

The low-energy spectrum of spin ladders is well studied
\cite{barne93,gopal94,greve96,shelt96,oitma96b,sushk98,eder98,jurec00,trebs00,knett01b,schmi05b}. 
All excitations can be classified by the parity 
with respect to reflections about the centerline of the spin ladder: an odd 
number of triplons has odd parity and  an even number of triplons  has even 
parity. 

The most relevant low-energy excitation has odd parity corresponding to a 
single triplon. It is completely characterized by the 
one-triplon dispersion
\begin {equation}
 \omega(k)=\nu_0+2\sum_{m=1}  \nu_m \cos(m k),
\label{dispersion1}
\end {equation}
where $k$ is the momentum along the ladder and the $\nu_m$ are the 
one-triplon hopping amplitudes. The dispersion takes a minimum at $k=\pi$ 
defining the one-triplon gap $\Delta$. In this work we are solely interested 
in one-triplon energies. Therefore, we do not discuss channels with more 
triplons. But we note that there are interesting interaction effects present 
in two-leg ladders leading to two-triplon bound and anti-bound states as well 
as pronounced continua in their response functions
\cite{uhrig96b,uhrig96be,shelt96,sushk98,damle98,trebs00,jurec00,knett01b,schmi05b}.

The dynamic structure factor $S^{\alpha,\beta}_T (k,\omega)$ is the quantity 
of interest for inelastic neutron scattering experiments. It is given by
\begin{equation}
 S^{\alpha,\beta}_T (k,\omega) = \frac{1}{1-e^{-\frac{\omega(k)}{T}}} 
\,{\rm Im} \, \chi^{\alpha,\beta}_T (k,\omega) \quad,
\end{equation}
where
\begin{equation}
 \chi^{\alpha \beta}_T (k,\omega)= i\int_{-\infty}^\infty e^{i \omega t} 
\sum_r e^{-ik r} \langle [ S_r^\alpha (t), S_0^\beta(0) ]\rangle \theta (t)
\end{equation}
denotes the retarded dynamic susceptibility and the superscripts
$\alpha,\beta\in\{\pm 1, 0 \}$ correspond to 
the components of the spin. Since our model is SU(2) invariant, the only 
finite components of the dynamic structure factor are the diagonal ones 
($\alpha=\beta$). We define 
\begin{equation}
 S_T (k,\omega)=S^{0,0}_T (k,\omega).
\end{equation}

The dynamic structure factor at zero temperature is dominated by the 
one-triplon contribution \cite{knett01b,schmi05b}. 
Theoretically, the general form of the one-triplon 
contribution (or more generically of any single mode (SM) approximation) reads 
\begin{equation}
  S_{T=0}^{\rm SM}(k, \omega)=a^2(k) \delta(\omega-\omega(k)),
  \label{SMA}
\end{equation}
where the prefactor $a^2(k)$ defines the one-triplon spectral weight which is 
known to be strongly peaked at the gap momentum $k=\pi$ 
\cite{uhrig04a,schmi05b}.

\section{Approach}
\label{A}

In this work we are interested in the effects of thermal fluctuations on the 
one-triplon contribution to the dynamic structure factor of the spin ladder. 
Physically, one expects that the triplon acquires a finite life-time upon 
heating. Then the $\delta$-functions at zero temperature are replaced by 
resonances which are found to show asymmetric line-shapes \cite{essle08}.    

Our approach is complementary. We do not describe the decay of 
triplons due to scattering with other thermally
excited triplons, but we develop an effective single-mode approximation.
That means that the one-triplon contribution to the dynamic structure factor 
is still given by a $\delta$-function, but with temperature-dependent spectral 
weight. On the level of the triplon dynamics, this is less sophisticated than 
the approach by Essler and collaborators \cite{james08,essle08,james09}. In 
contrast, we are able to efficiently incorporate vertex corrections which are 
also relevant at finite temperatures. This leads to a 
temperature-dependent one-triplon spectral weight in our theory. 

In the following we give a very short introduction to the method of 
continuous unitary transformations focusing on
 the perturbative realization which is used in this work.
The acronym CUT stands for the continuous unitary transformations in general 
and PCUT for their perturbative realization in particular.

We concentrate on the vertex corrections which are relevant for the single-mode
 description of the finite-temperature dynamic structure factor. For a 
detailed introduction to CUTs we refer to the literature, see below. In the 
sequel, we describe the mean-field decoupling of the vertex corrections in 
detail. They give access to the effective temperature-dependent description 
of the dynamic structure factor at low energies. 

 \subsection{(P)CUTs}
The CUT method was invented independently by Wegner \cite{wegne94} and 
G\l{}azek  and Wilson \cite{glaze93,glaze94} in 1994. 
The CUT itself is defined through the flow equation 
\begin{equation}
 \partial_l H(l)=\left[ \eta (l), H(l)\right]
\label{defcut}
\end{equation}
with the continuous parameter $l \in [0, \infty ]$ and the initial condition 
$H(l=0)=H$ where $H$ is the given Hamiltonian. The final form of the effective 
Hamiltonian $H_{\text{eff}}=H(l=\infty )$ depends  on the choice of the 
infinitesimal generator $\eta$. 

Here we use (quasi)particle conserving CUTs 
\cite{mielk98,uhrig98c,knett00a,fisch10a} 
which result in the effective Hamiltonian $H_{\text{eff}}$ being 
block-diagonal in the number of elementary excitations $q$. In the
case of the spin ladder studied, $q$ is the number of excited triplons.
In an eigenbasis  $|q_i\rangle$ ($i$ is an additional index necessary due
to degeneracies) of 
the particle-number operator $Q$ with eigenvalues $q\in\mathbb{N}_0$, the 
generator $\eta^{\rm pc}$ is given by  
\begin{equation}
\label{generator-pc}
 \eta^{\rm pc}_{ij}(l)=\mathrm{sgn}(q_i-q_l)H_{ij}(l).
\end{equation}
For the two-leg ladder we choose $J_\perp Q=H_\perp$ giving an effective 
Hamiltonian which conserves the number of triplons, i.e., 
$\left[ H_\perp, H_{\text{eff}}\right]=0 $. Consequently, the effective 
Hamiltonian $H_\text{eff}$ decomposes into a sum of \emph{irreducible} 
$n$-triplon operators $H_n$
\begin{equation}
 H_\text{eff}=\sum_{n=0}^\infty H_n,
\label{Hn}
\end{equation}
and each $n$-triplon block can be treated separately, see Refs.\ 
\onlinecite{knett03a,fisch10a} for details.

We apply the CUTs in their perturbative version, i.e., the effective 
Hamiltonian for the two-leg ladder is obtained as a high-order series 
expansion in the small  parameters 
\begin{subequations}
\begin{eqnarray}
x &:=& J_\parallel/J_\perp\\
x_{\rm cyc} &:=& J_{\rm cyc}/J_\perp.
\end{eqnarray}
\end{subequations}
The essential one-triplon dispersion $\omega(k)$
has already been determined earlier  \cite{schmi05b,knett03a}. The 
one-triplon hopping amplitudes $\nu_m$ have been calculated up to order 11 in 
both perturbative parameters. Various extrapolation schemes 
can be applied successfully \cite{schmi03d}. Especially the low-energy part 
of the dispersion including the one-triplon gap can be determined very 
reliably up to realistic values $x\approx 1.25-1.5$ and 
$x_{\rm cyc}\approx 0.2-0.25$.

In order to calculate spectral properties of the system such as the dynamic 
structure factor $S_T(k, \omega)$ one has to transform observables $O$ by the
same CUT as the Hamiltonian. Hence the same generator $\eta^{\rm pc}$
given in Eq.\ \eqref{generator-pc} is used
\begin{equation}
  \partial_l O(l)=\left[ \eta^{\rm pc}(l), O(l)\right].
\label{defcut2}
\end{equation}
The relevant observable $O(r)$ with odd parity for the dynamic structure 
factor \cite{schmi05b} is given locally on rung $r$ by
\begin{equation}
\label{observ0}
 O(r) =S_{r,1}^{z}-S_{r,2}^{z}=t_{0,r}^\dagger
s_{0,r}^{\phantom{\dagger}} + t_{0,r}^{\phantom{\dagger}}s_{0,r}^\dagger.
\end{equation}
The bare observable  creates or destroys a single triplon with flavor 
$0$ on rung $r$. After the CUT, the effective observable $O^{\rm eff}(r)$ 
comprises many channels reflecting the complicated nature of the interacting 
triplon problem. One finds  
\begin{equation}
 O^{\rm eff}(r) =U^\dagger O (r)  U=\sum_{n,m}  O_{n,m}^{\text{eff}}(r) 
\end{equation}
where $O_{n,m}^{\text{eff}}(r)$ stands for a process
where $m$ triplons are annihilated and $n$ are created.
Thus $n,m \in \mathbb{N}_0$ with the constraint that the difference $n-m$ is 
an odd number reflecting the odd parity of the observable. At zero
temperature, no triplons are present so that only 
the terms  $O_{n,0}$ with $n$ odd are relevant.

Below we need the most relevant one-triplon contribution $O_{1,0}$ which 
is defined by the one-triplon spectral weights 
$a^2 (k)$. The corresponding real space amplitudes have been calculated 
earlier up to order 10 \cite{schmi05b}. Extrapolations of the low-energy part 
close to $k=\pi$ are reliable up to realistic values of $x$ and  $x_{\rm cyc}$.
The contributions $O_{n,0}$ with $n\ge 3$ are significantly smaller
in weight than   $O_{1,0}$ \cite{knett01b,schmi05b}.

At finite temperatures, also contributions $O_{n,m}$ with $m>0$ matter. These 
contributions give rise to vertex corrections. We will argue below that the 
most relevant vertex correction for the finite temperature physics at low 
energies of the spin ladder is the contribution $O_{2,1}$. It comprises
processes where a second triplon is created assuming another triplon is already
 present in the system due to thermal fluctuations. Thus we refer
to these processes as conditional excitation processes.

We determined this contribution up to order 9 in the perturbative 
parameters $x$ and $x_\text{cyc}$. 
For extrapolation, we used the method of internal variables 
\cite{schmi03a}. Unfortunately, no reliable Pad\'e or DlogPad\'e 
resummation on top of this extrapolation scheme succeeded, in contrast to
the extrapolation of the 
one-triplon dispersion and the one-triplon spectral weights. Certainly, this
fact reflects the more complicated nature of conditional excitations. 
Nevertheless, we expect that the extrapolation still yields 
reasonable values for the physical processes 
up to realistic values of $x$ and $x_{\rm cyc}$.

\subsection{Effective single-mode approximation}
\label{StarPoi}

Next we set up our temperature-dependent theory which  results in an 
effective single-mode approximation. A single-mode contribution to the 
dynamic structure factor (see Eq.~(\ref{SMA})) depends solely on the 
dispersion and on the spectral weight. The general expression reads 
\begin{equation}
 S_{T}^{\rm SM}(k, \omega)=\frac{1}{1-e^{-\frac{\omega_T (k)}{T}}} \, a_T^2(k) 
\, \delta(\omega-\omega_T(k)),
\label{SMAT}
\end{equation}
where we set the Boltzmann constant $k_\text{B}$ to unity, i.e.,
we measure the temperature in units of energy henceforth.

First, we discuss the temperature effects in the effective
dispersion $\omega_T(k)$. Second, we  incorporate vertex corrections
in the spectral weights $a_T^2(k)$.

\subsubsection{Effective dispersion}

The one-triplon contribution is characterized at zero 
temperature by $\omega(k)$. In the PCUT calculation the fully-condensed 
singlet state $|{\rm ref}\rangle$ with $\langle s^\dagger s \rangle =1 $ is 
used as the reference state. Hence no parameter $s < 1$ for the
singlet occupation needs to be introduced as it was used previously
to take the hardcore constraint into account \cite{sachd90,gopal94}.
The CUT takes the hardcore constraint into account.

At finite temperatures, however, the situation is different. Even after
the CUT there will be a finite occupation of triplons which lowers the
singlet occupation. In analogy to the previous
studies at zero temperature, we take this effect into account
by introducing  the mean-field parameter $s(T)$  
defined as the condensate value $\langle s^\dagger \rangle =s$
 with $s\in [0,1 ]$. We stress the difference to previous studies
\cite{sachd90,gopal94} where $s<1$ was used to consider quantum fluctuations
while we use this concept only to consider thermal fluctations.
More complex expressions such as $s_j^\dag s_i^{\phantom\dag}$ are
reduced to $s^2$ which focusses  on the average value 
neglecting fluctuations.

Then the $s$-dependent effective one-triplon dispersion reads
\begin {equation}
 \omega_T(k,s)=\nu_0+2s^2 (T) \sum_{m=1}  \nu_m \cos(m k)
\label{dispersion2}
\end {equation}
because each non-local hopping 
$t_{\alpha,j}^\dag t_{\alpha,i}^{\phantom\dag}$  
with $j\ne i$ implies the transfer of a singlet 
$s_{i}^\dag s_{j}^{\phantom\dag}$. The local processes 
$t_{\alpha,j}^\dag t_{\alpha,j}^{\phantom\dag}$  do not require
such a renormalization.

To determine  the temperature-dependent mean field parameter 
$s(T)$ it is inserted in the hardcore 
constraint Eq.~(\ref{bedingung}) yielding
\begin{equation}
 1=s^2(T)+3 \left\langle t^\dagger t \right\rangle(T).
\label{norm}
\end{equation}
So the average triplon occupation 
$\left\langle t^\dagger t \right\rangle(T)$ determines $s(T)$.
Due to the spin symmetry the expectation values $\left\langle t_\alpha^\dagger
 t_\alpha^{\phantom{\dagger}} \right\rangle (T)$ with $\alpha \in \left\lbrace 
-1, 0, 1  \right\rbrace $ are independent of $\alpha$ resulting in the factor 
3 in Eq.\ \eqref{norm}. 

The average triplon occupation  can be calculated by the sum over all modes 
$\langle t_k^\dagger t_k^{\phantom{\dagger}} \rangle$ in momentum space. 
At finite temperatures one has
\begin{equation}
 \left\langle t^\dagger_k t^{\phantom{\dagger}}_k \right\rangle=\frac{1}{Z} 
\mathrm{Tr}\left(t^\dagger_k t^{\phantom{\dagger}}_k e^{-\frac{H}{T}} \right)=
\frac{\partial F}{\partial \omega_k}  
\label{mittelwert}
\end{equation}
where $k$ is the triplon momentum. The free energy $F$ is defined via the 
partition function $Z$ by
\begin{equation}
 F= -T \ln(Z)
\label{freieenergie}
\end{equation}
where $k_\textrm{B}$ is set to unity.
Integration over the whole Brillouin zone yields the wanted local 
triplon density
\begin{equation}
 \left\langle t^\dagger t \right\rangle\left( T\right)  =\int_{-\pi}^\pi 
\frac{\left\langle t^\dagger_k t^{\phantom{\dagger}}_k \right\rangle} 
{2 \pi}dk.
\label{mittel}
\end{equation}

If the triplons were bosons without contraint 
$\left\langle t^\dagger_k t^{\phantom{\dagger}}_k \right\rangle$
would be given by the standard bosonic occupation $(\exp(\omega(k)/T)-1)^{-1}$.
Unfortunately, there is no equivalent rigorous expression for hardcore bosons.
As an approximate treatment we follow the arguments of Troyer et al.\
\cite{troye94} to reweight the partition function $Z$ yielding 
\begin{subequations}
\begin{eqnarray*}
 Z &=& \left(Z_\omega\right)^N
\\
Z_\omega &:=& 1+\frac{3}{2\pi}\int_{-\pi}^\pi e^{- \frac{ \omega_T(k,s)}{ T}}dk
\end{eqnarray*}
\label{partition}
\end{subequations}
where $N$ is the number of rungs. Consequently, the free energy becomes
\begin{equation}
F 
=-N T\ln\left(1+\frac{3}{2\pi}\int_{-\pi}^\pi e^{-\frac{\omega_T(k,s)}{T}}dk
\right).
\label{freegie}
\end{equation}
Combining Eqs.\ (\ref{freegie},\ref{mittel},\ref{norm}) yields
\begin{equation}
 s^2(T)=1-\frac{3}{2 \pi}\frac{1}{Z_{\omega}}\int_{-\pi}^\pi  
e^{-\frac{\omega_T(k,s)}{T}}dk.
\label{result1}
\end{equation}
Due to the dependence of the dispersion $\omega_T(k,s)$ on $s$
the above equation defines $s$ only implicitly. It has to be evaluated
 self-consistently. 

The value of $s^2$ lies in the interval $[0,1]$.
Clearly $\lim_{T\to 0} e^{-\frac{\omega_T(k,s)}{T}} =0$
as long as $\omega_T(k,s)\ge C >0$ which is the physically
reasonable case. Hence $s^2(T=0)=0$ follows. Similarly, the
limit $T\to\infty$ is given by 
$\lim_{T\to \infty}e^{-\frac{\omega_T(k,s)}{T}}=1$
so that $s^2(T=\infty)=1/4$ is implied.

For completeness, we mention another effect
which implies a renormalization of the effective dispersion
due to a finite concentration of thermally excited triplons.
Triplon-triplon interactions of the type 
$t^\dagger_1 t^\dagger_2 t^{\phantom\dag}_3t^{\phantom\dag}_4$ 
are known to exist. For instance, they lead to 
the occurrence of bound states \cite{windt01,knett01b}.
They also yield some renormalization of the dispersion upon
mean-field decoupling 
\begin{eqnarray}
t^\dagger_1 t^\dagger_2 t^{\phantom\dag}_3t^{\phantom\dag}_4 &\approx& 
\langle t^\dagger_1 t^{\phantom\dag}_3\rangle t^\dagger_2 t^{\phantom\dag}_4 
+
\langle t^\dagger_2 t^{\phantom\dag}_3\rangle t^\dagger_1 t^{\phantom\dag}_4
\nonumber\\ 
\label{inter-MF}
&& +
\langle t^\dagger_1 t^{\phantom\dag}_4\rangle t^\dagger_2 t^{\phantom\dag}_3
+
\langle t^\dagger_2 t^{\phantom\dag}_4\rangle t^\dagger_1 t^{\phantom\dag}_3
+\text{const.},
\qquad 
\end{eqnarray}
which is proportional to the triplon densities
$\left\langle t^\dagger t \right\rangle(T)$.
The inclusion of the above terms will renormalize
the effective dispersion $\omega_T(k)$ slightly.
We do not consider the effect of the
terms in Eq.\ \eqref{inter-MF} here quantitatively for two reasons. 

First, an estimate of the quantitative effect
of these corrections indicates that they are not very significant.
For instance, they are of the same order as the effect of the
inclusion of the factor $s^2(T)$ in Eq.\ \eqref{dispersion2}.
If this term is omitted no qualitative changes of our results
will occur because they are dominated by the renormalization
of matrix elements, not by the renormalization of the energies
as we will show in the following. 
Second, their systematic treatment
 is very tedious because a quantitative determination
of all interaction elements would be required.

\subsubsection{Vertex corrections and effective spectral weights}
\label{SpinOp}

\begin{figure}
  \includegraphics[scale=0.4]{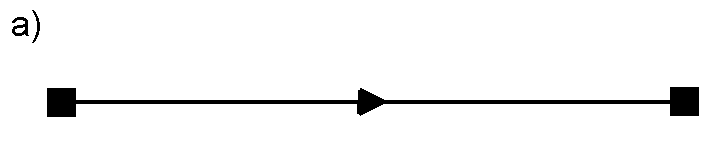}\\
  \includegraphics[scale=0.4]{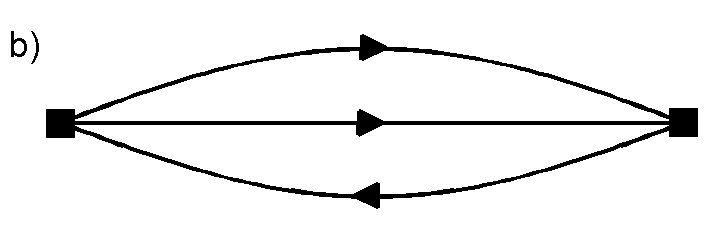}
  \caption{Propagation of the excitation  included in the calculation. 
    Diagram a) shows the \emph{one triplon excitation}; diagram b)
    shows the \emph{conditional triplon excitation} which requires
    the existence of a triplon.}
  \label{condex}
\end{figure}

The second building block for the one-triplon contribution to the dynamic 
structure factor is the one-triplon spectral weight.  Its unconditional part,
active also at zero temperature, is captured by $O_{1,0}^{\text{eff}}$  after 
the CUT. Its propagation is illustrated diagrammatically in panel a) of 
Fig.\ \ref{condex}.
Contributions $O_{n,m}^{\text{eff}}$ with $m>0$ are 
only active at finite temperatures; they represent conditional excitations.
Our idea is to include such contributions 
as vertex corrections in an effective manner. We identify 
$O_{2,1}^{\text{eff}}$ as most important because it requires
only a single triplon to be present. Hence its quantitative contribution
will be proportional to the triplon density.
In second quantization  $O_{2,1}^{\text{eff}}$  contains operators of the form 
$t^\dagger_\alpha t^\dagger_\beta t^{\phantom{\dagger}}_\gamma$. 
It is illustrated diagrammatically in panel b) of Fig.\ \ref{condex}.

Any term of $O_{m+1,m}^{\text{eff}}$ with $m>1$ requires the
existence of $m$ triplons so that its quantitative contribution will
be proportional  to the $m$-th power of the triplon density. In the regime
of low temperatures not  larger than about the spin gap each
additional power in the triplon density stands for another additional
exponentially small factor.

\begin{figure}[ht]
  \centering
  \includegraphics[scale=0.45]{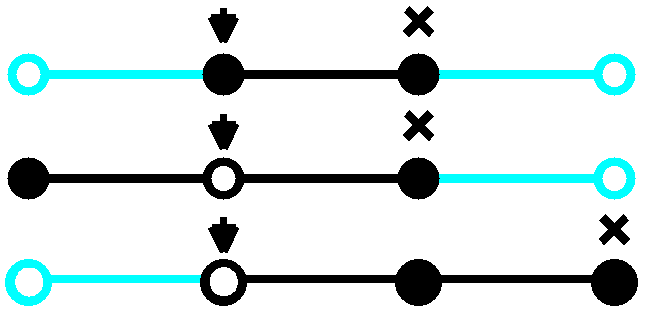}
  \caption{(color online) Some conditional excitation processes are sketched 
    which are of low order in $x$. The arrows indicate the rung on which 
    the observable $O_{2,1}^{\text{eff}}(r)$ acts, whereas the cross 
    \textbf{x} denotes the rung which is occupied by the 
    one-triplon before the conditional excitation. Black filled circles denote 
    the rungs occupied by triplons after the conditional excitation process. 
    Open circles are in their singlet state after the excitation; the cyan 
    (gray) part is shown for completeness.}
  \label{obser}
\end{figure}

Restricting to the two parts  $O_{m+1,m}^{\text{eff}}$
with $m\in\{ 0,1\}$, the effective observable is given by
\begin{eqnarray}
  O^{\rm eff} (r)  &=& U^\dagger O(r) U
  \nonumber\\
  &\approx& O_{1,0}^{\text{eff}}(r) + O_{2,1}^{\text{eff}}(r)
  \nonumber\\
  &= &\sum_p \left( a^{\phantom{\dagger}}_p t^\dagger_{0,r+p}+  \textrm{h.c.}
  \right) 
  \nonumber\\
  && \hspace*{-1cm}
  +\sum_{\alpha, \beta, \gamma} \sum_{i<j,p} \left( a_{i,j,p}^{\alpha, \beta, 
    \gamma} t^\dagger_{\alpha, r+i} t^\dagger_{\beta,r+j} 
  t^{\phantom{\dagger}}_{\gamma,r+p} +  \textrm{h.c.}\right).\quad
  \label{Sl}
\end{eqnarray}
The ordering of operators in the above equation is not unique. 
To avoid double counting,  the convention $i < j$ is used. 
A few excitation processes appearing in 
$O_{2,1}^{\text{eff}}$ in low order in the small parameters are illustrated in 
Fig.~\ref{obser}. 

Symmetries of the system imply further constraints.
The conservation of the total $S^z$ component for processes of the type 
$t^\dagger_\alpha t^\dagger_\beta t_\gamma$ lead to
\begin{equation}
 0=\alpha+\beta-\gamma \quad \quad \alpha, \beta,\gamma \in \left\lbrace 0,  
\pm 1\right\rbrace ,
\label{constraint}
\end{equation}
where it is understood that the bare observable is given by 
Eq.\ \eqref{observ0}.

For scattering experiments we are interested in the Fourier transform
of $ O^{\rm eff} (r)$ in Eq.~(\ref{Sl}). For the unconditional
contribution $O_{1,0}^{\text{eff}}$ we find
\begin{equation}
\begin{split}
O_{1,0}^{\text{eff}}(k) &= \frac{1}{\sqrt{N}} \sum_{r}  e^{\mathrm i k r}  
O_{1,0}^{z, \text{eff}}(r) 
\\
&= \frac{1}{\sqrt{N}} \sum_{r,p} \left( e^{ \mathrm i k r } 
a^{\phantom{\dagger}}_p t^\dagger_{0,r+p}+ e^{\mathrm i k r } 
a^{\phantom{\dagger}}_p t^{\phantom{\dagger}}_{0,r+p}\right) 
\\
&= \frac{1}{\sqrt{N}}\sum_{r,p} a^{\phantom{\dagger}}_p\left(  
e^{ \mathrm i k (r-p)} t_{0,r}^\dagger+e^{ \mathrm i k (r-p)} 
t^{\phantom{\dagger}}_{0,r} \right) 
\\
&=  a(k) \left(  t_{0, -k}^\dagger+  t^{\phantom{\dagger}}_{0, k} \right)
\end{split}
\label{rq}
\end{equation}
where we use
\begin{equation}
a(k) := \sum_pe^{ -\mathrm i k p} a_p.
\end{equation}
Note that the coefficients $a_p$ are real. For the symmetric spin ladder
studied here they are also symmetric $a_p=a_{-p}$.
The operators $t_{0, k}^\dagger$ and $t_{0, k}$ are defined by
\begin{subequations}
\label{operatordef}
\begin{eqnarray}
 t_{0, k}^{\phantom\dagger} &:=& \frac{1}{\sqrt{N}}\sum_r
e^{ \mathrm i k r}t^{\phantom\dagger} _{0, r}
 \\
 t_{0, k}^{\dagger} &:=& \frac{1}{\sqrt{N}}\sum_r
e^{ -\mathrm i k r}t^{\dagger} _{0, r}.
\end{eqnarray}
\end{subequations}

The complete treatment of the conditional excitation 
is complicated since it involves several sums
and the resulting terms are not diagonal in momentum space. 
We do not, however, aim at the exhaustive description of the multi-particle
response, but at the renormalization of the single-mode
response. Therefore, we use a mean-field decoupling 
for $O_{2,1}^{\text{eff}}(k)$ to identify the processes belonging
to the creation of a single triplon. 

\begin{figure}[htbp!]
  \centering
  \includegraphics[scale=0.18]{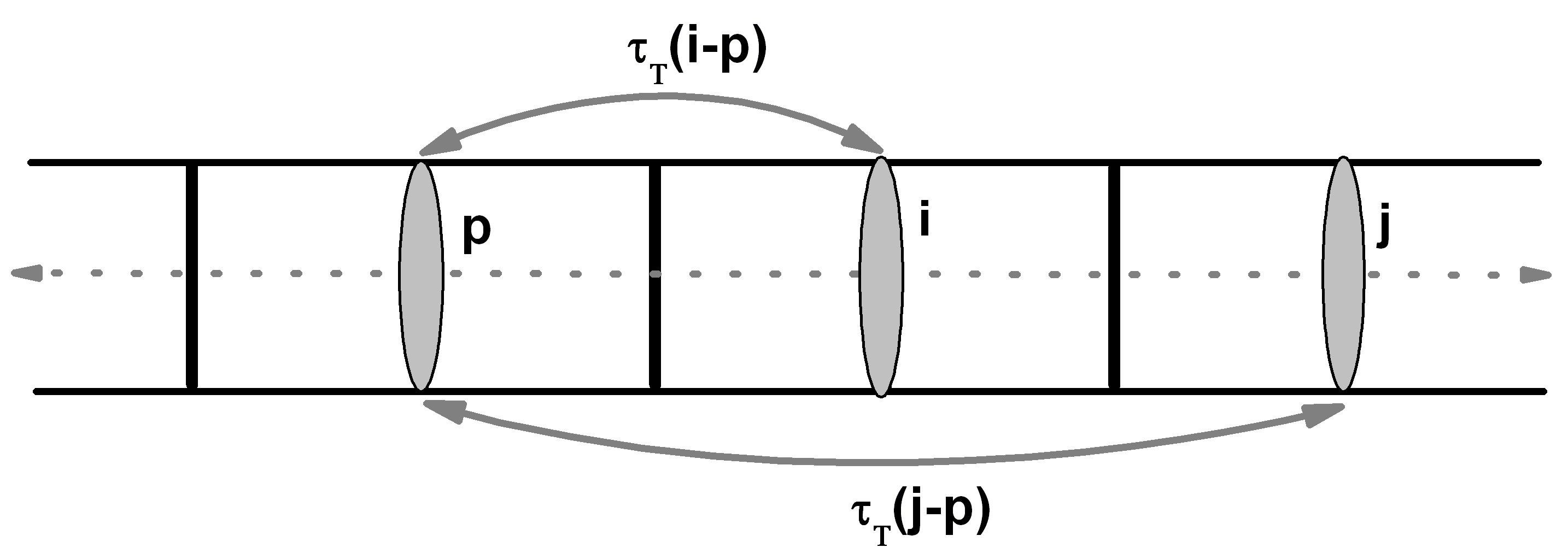}
  \caption{Illustration of the hopping expectation value 
    $\tau_T(j-p)$ as defined in Eq.~\eqref{hop-exp-val}.
  }
  \label{leiter}
\end{figure}

The mean-field decoupling is
motivated by Wick's theorem \cite{wick50}, but represents an approximation
in the present context
because the triplons are hardcore bosons. For concentrations of the triplons
tending to zero, i.e.,
for $T\to 0$, the hardcore constraint becomes ineffective and
the use of Wick's theorem is justified
\begin{equation}
  t^\dagger_{\alpha,i} t^\dagger_{\beta,j} t^{\phantom{\dagger}}_{\gamma,p} 
  \approx t^\dagger_{\alpha, i} \langle t^\dagger_{\beta,j} 
  t^{\phantom{\dagger}}_{\gamma,p} \rangle + t^\dagger_{\beta, j} 
  \langle t^\dagger_{\alpha,i} t^{\phantom{\dagger}}_{\gamma,p} \rangle.
  \label{decoupling}
\end{equation}
Due to the symmetry constraint (\ref{constraint}) the expectation values
are non-zero only for the cases
\begin{equation}
  \begin{split}
    \langle t^\dagger_{\beta,j} t^{\phantom{\dagger}}_{\gamma,p} 
    \rangle \neq 0 &\quad\text{if}\quad \beta=\gamma \quad\text{and}
    \quad \alpha=0
    \\
    \langle t^\dagger_{\alpha,i} t^{\phantom{\dagger}}_{\gamma,p} 
    \rangle\neq 0 
    &\quad \text{if}\quad\alpha=\gamma \quad \text{and} \quad \beta=0.
  \end{split}
\end{equation}
In all other cases the expectation values are zero.
We define the \emph{hopping expectation values} by
\begin{equation}
  \label{hop-exp-val}
  \tau^{\phantom{\dagger}}_T (j-p) := \langle t^\dagger_{\beta,j} 
  t^{\phantom{\dagger}}_{\beta,p} \rangle.
\end{equation}
Due to spin rotation symmetry and translation symmetry
this definition is independent of $\beta$,
but depends only on the relative distance of the two rungs $j$ and $p$  as 
depicted in Fig.~\ref{leiter}. Consequently, we perform an index shift 
$(j-p) \rightarrow j$ yielding $\tau_T(j)$ which is used in the following.

Employing the mean-field decoupling \eqref{decoupling} renders the
operator structure of $O_{2,1}^{\text{eff}}$ the same as for 
$O_{1,0}^{\text{eff}}$. Therefore,  Fourier transformation makes it
concise
\begin{eqnarray}
O_{2,1}^{\text{eff}}(k)&=& \frac{1}{\sqrt{N}}\sum_{r, i,j,p} \sum_{\alpha, 
\beta, \gamma} a_{i,j,p}^{\alpha, \beta, \gamma}  \left( e^{ \mathrm i k r}
  t^\dagger_{\alpha, r+i} t^\dagger_{\beta,r+j} 
t^{\phantom{\dagger}}_{\gamma,r+p} \right. 
\nonumber\\
&&\left. +  e^{\mathrm i k r} t^{\phantom{\dagger}}_{\alpha, r+i} 
t^{\phantom{\dagger}}_{\beta,r+j} t^\dagger_{\gamma,r+p} \right) 
\\
&\approx & \frac{1}{\sqrt{N}} \sum_{r, i,j,p}  \sum_{ \beta}  
a_{i,j,p}^{0, \beta, \beta}  \left( e^{ \mathrm i k r}  t^\dagger_{0, r+i} 
\tau^{\phantom{\dagger}}_T(j-p) 
\delta^{\phantom{\dagger}}_{\beta, \gamma}\right.
\nonumber\\
&& \left.+ e^{\mathrm i k r} t^{\phantom{\dagger}}_{0, r+i} 
\tau^{\phantom{\dagger}}_T(j-p) \delta^{\phantom{\dagger}}_{\beta, \gamma} 
\right)  
\nonumber\\
&& + \frac{1}{\sqrt{N}} \sum_{r, i,j,p}\sum_\alpha 
a_{i,j,p}^{\alpha, 0, \alpha} \left( e^{ \mathrm i k r} t^\dagger_{0, r+j} 
\tau^{\phantom{\dagger}}_T(i-p) \delta^{\phantom{\dagger}}_{\alpha, \gamma} 
\right.
\nonumber\\
&&\left.+ e^{\mathrm i k r}  t^{\phantom{\dagger}}_{0, r+j} 
\tau^{\phantom{\dagger}}_T(i-p) 
\delta^{\phantom{\dagger}}_{\alpha, \gamma}\right). 
\end{eqnarray}
Shifting the indices $r \to r-i$ and $j\to j+p$ in the first sum as well as
$r \to r-j$ and $i\to i+p$ in the second sum leads to
\begin{eqnarray}
&&O_{2,1}^{\text{eff}} (k)\ =
\nonumber\\
&&\quad\frac{1}{\sqrt{N}} \sum_{r, i,j,p} 
\sum_\beta a_{i,j+p,p}^{0, \beta, \beta}  \left( e^{- \mathrm i k i}  
e^{ \mathrm i k r}t^\dagger_{0, r}
\right.
\nonumber\\
&&\ \left. 
+ e^{- \mathrm i k i}  e^{ \mathrm i k r} 
t^{\phantom{\dagger}}_{0, r}   \right)\tau^{\phantom{\dagger}}_T(j) 
\nonumber\\
&&\quad + \frac{1}{\sqrt{N}} \sum_{r, i,j,p}\sum_\alpha 
a_{i+p,j,p}^{\alpha, 0, \alpha} \left(e^{- \mathrm i k j}  
e^{ \mathrm i k r} t^\dagger_{0, r} 
\right.
\nonumber\\
&&\ \left. 
+e^{ -\mathrm i k j} e^{\mathrm i k r}  t_{0, r} \right) 
\tau^{\phantom{\dagger}}_T(i)
\\
&&\overset{(\ref{operatordef})}{=} \sum_{i,j,p} \sum_\beta 
a_{i,j+p,}^{0, \beta, \beta}  \left(   t^\dagger_{0, -k}  +   
t^{\phantom{\dagger}}_{0,k}   \right)e^{- \mathrm i k i}  
\tau^{\phantom{\dagger}}_T(j)
\nonumber\\
& &\ + \sum_{ i,j,p}\sum_\alpha a_{i+p,j,p}^{\alpha, 0, \alpha} 
\left(  t^\dagger_{0, -k} + t^{\phantom{\dagger}}_{0, k}\right) 
e^{ -\mathrm i k j} \tau^{\phantom{\dagger}}_T(i)
\\
&& = \sum_{ i,j,p}\sum_\alpha 
\left(a_{j,i+p,}^{0, \alpha, \alpha} +a_{i+p,j,p}^{\alpha, 0, \alpha} \right)
\left(  t^\dagger_{0, -k} + t^{\phantom{\dagger}}_{0, k}\right) 
e^{ -\mathrm i k j} \tau^{\phantom{\dagger}}_T(i).
\nonumber
\end{eqnarray}

Finally the full effective observable $O^{\rm eff} (k)$ in momentum 
space is  given by
\begin{eqnarray}
 O^{\rm eff} (k) &=& O_{1,0}^{\text{eff}}(k)+O_{2,1}^{\text{eff}}(k)
\nonumber \\
&=&  \left[a(k) + A_T(k)\right] 
\left(  t_{0, -k}^\dagger+ t^{\phantom{\dagger}}_{0, k} \right), 
\label{spec-weight-renorm}
\end{eqnarray}
where we used the successive definitions
\begin{subequations}
  \label{Adefs}
  \begin{eqnarray}
    A_T(k) &:=& \sum_i A_i(k) \tau_T(i)
    \\
    A_i(k) &:=& \sum_j A_{i,j} e^{-\mathrm i kj}\\
    A_{i,j}  &:=& \sum_{p,\alpha} 
    \left(a_{j,i+p,}^{0, \alpha, \alpha} +a_{i+p,j,p}^{\alpha, 0, \alpha} 
    \right).
  \end{eqnarray}
\end{subequations}
Clearly, \eqref{spec-weight-renorm} shows that the conditional excitations
imply a renormalization of the spectral weight of the triplons
which are the single-mode excitations for the spin ladder.
The temperature dependence of $ A_T(k)$ stems from the temperature
dependence of the hopping expectation values $\tau_T(j)$. Similar to the
calculation of the triplon occupation in \eqref{mittel}, the hopping
expectation values result from the suitably weighted
 averages over the Brillouin zone
 \begin{equation}
   \label{nj}
   \tau_T(j)  = \frac{1}{2 \pi Z_\omega}\int_{-\pi}^\pi 
   e^{-\frac{\omega_T(k)}{T}} \cos (j k) dk.
 \end{equation}
 
The expressions \eqref{Adefs} for the effective spectral weights hold for 
all momenta $k$. 
But the main effect in the spectral weight is measured at the lowest excitation
energy $\Delta:=\omega(\pi)$, i.e., at the momentum $k=\pi$. 
Hence we restrict ourselves to $k=\pi$ in the following.
For $A_i(\pi)$ we obtain explicitly
\begin{eqnarray}
  A_i(\pi) &=& \sum_{p, j}(-1)^j \left(  a_{j,i+p,p}^{0, 0, 0}+
  a_{i+p,j,p}^{0, 0, 0} \right.
  \nonumber \\
  &&\left. + 2 \left( a_{j,i+p,p}^{0, 1, 1}+a_{i+p,j,p}^{1, 0, 1} 
  \right)\right),
  \label{grossa}
\end{eqnarray}
where the factor $2$ comes from the fact that 
$a_{j,i+p,p}^{0, 1, 1}=a_{j,i+p,p}^{0, -1, -1}$ and 
$ a_{j,i+p,p}^{1, 0, 1}=a_{j,i+p,p}^{-1, 0, -1}$, respectively.

Adding the unconditional and the conditional excitation processes yields  the 
total effective temperature-dependent one-triplon spectral weight at $k=\pi$ 
\begin{equation}
 a^2_T (\pi) = \big[ a(\pi) + A^{\phantom{\dagger}}_T (\pi) \big]^2 s^2(T).
\label{transition}
\end{equation}
The factor $s^2(T)$ stands for the reduced availability of rungs for the
creation of triplons if there are already some thermally excited.
Putting all pieces together we generalize \eqref{SMA} to finite
temperatures
\begin{equation}
  S_{T}^{\rm SM}(k, \omega)=a^2_T(k) \delta(\omega-\omega_T(k)),
  \label{SMA-vertex}
\end{equation}
which we use in the present study for $k=\pi$ only.
This is the dynamic structure factor in
single-mode approximation of the dynamic structure factor including
the vertex corrections relevant at finite temperatures. 
The physical content of  Eq.\ \eqref{SMA-vertex} will be exploited in the 
next section.

\section{Theoretical Results}
\label{TR}

Next we  present our theoretical results obtained by extending the PCUT 
result to finite temperature using the mean-field (MF) approach. 
We demonstrate that the 
intensity decrease in  the dynamic structure factor mainly arises from the 
contribution of the conditional excitations.
 The basic energy unit in this theory section is chosen to be the rung
 coupling constant $J_\perp$ allowing for direct comparison  
for different parameter sets $\left\lbrace x, x_{\rm cyc}\right\rbrace $. 

\begin{figure}[htpb!]
  \centering
  \includegraphics[scale=0.29]{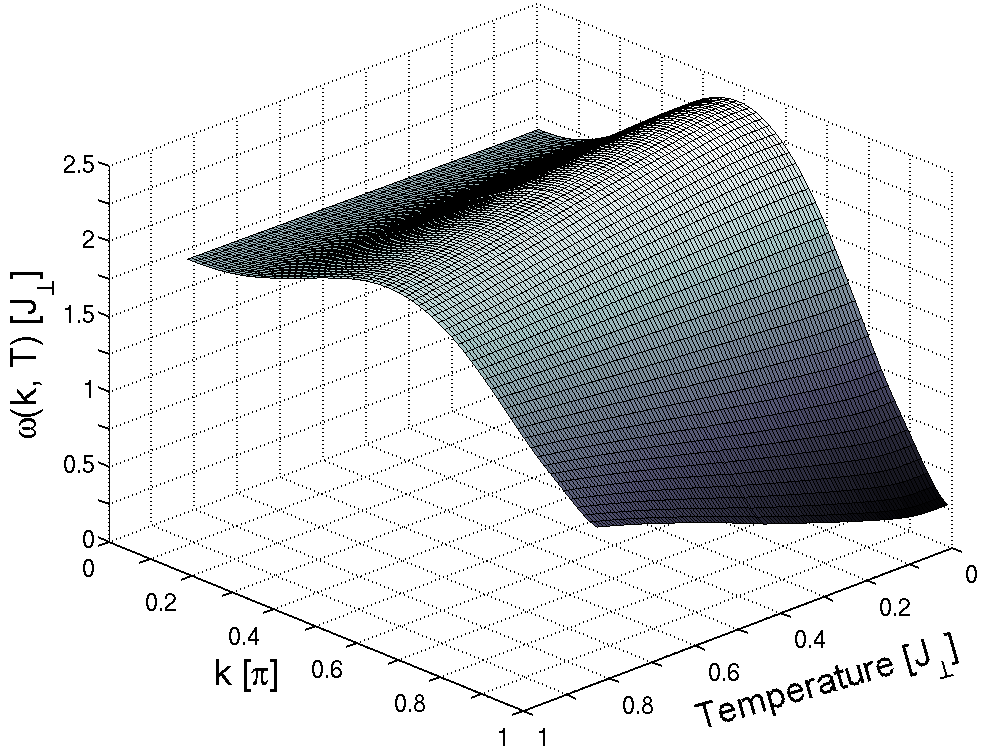}
  \caption{Effective mean-field dispersion  for 
    $x_{\rm cyc}=0.2$ and $x=1.5$  in dependence on temperature and momentum.}
  \label{surf}
\end{figure}

In Ref.\ \onlinecite{notbo07} the parameters $x=1.5$ and $x_{\rm cyc}=0.2-0.25$
were determined to describe a generic cuprate spin ladder best.
Thus we chose these parameters to compute the dependence of 
the self-consistent MF dispersions
shown in Fig.\ \ref{surf} with temperature $T$ and wave vector $k$.
This  provides information on how  the shape of the effective 
dispersion changes upon increasing temperature and momentum.
The dispersion becomes flatter on increasing temperature
implying a larger energy gap  $\Delta(T)=\omega(\pi,T)$.

\begin{figure}[htpb]
  \centering
  \includegraphics[scale=0.6]{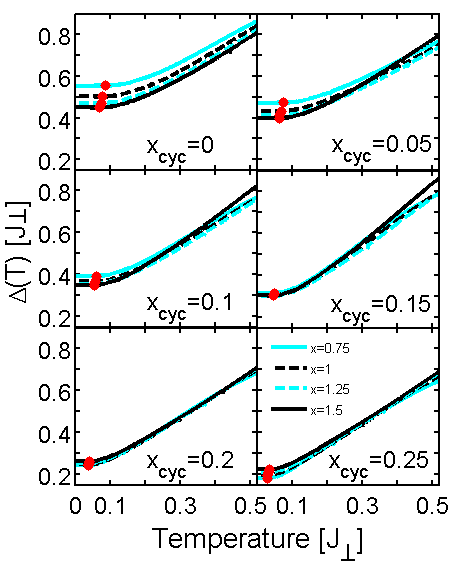}
  \caption{(color online) Excitation gap vs.\ temperature $T$  
    for $x=\left\lbrace  0.75, 1, 1.25, 1.5\right\rbrace $ and a 
    different $x_{\rm cyc}$ in each panel. The red (gray) dots indicate the 
    position of the characteristic temperature $T_{\text{char}}$.}
  \label{gapT}
\end{figure}
Figure \ref{gapT} shows the gap as a function of temperature 
for various $x=\left\lbrace  0.75, 1, 1.25, 1.5\right\rbrace$ 
and a different $x_{\rm cyc}$ in each panel. The blue (gray) 
dots in each panel 
indicate a characteristic temperature $T_{\text{char}}$ above which the spin
gap shows a significant dependency on temperature in form of a steep gap 
function $\Delta(T)$ for all $x$ and $x_{\rm cyc}$.
The value of $T_{\text{char}}$ scales with the spin gap $\Delta(T=0)$;
hence it decreases with increasing $x_{\rm cyc}$.

Three further striking points are to be mentioned. First, with increasing 
$x_{\rm cyc}$ the curves for $x=\left\lbrace  0.75, 1, 1.25, 1.5\right\rbrace$
approach one another until they lie almost on top of one another for  
$x_{\rm cyc}=0.15-0.2$. For  $x_{\rm cyc}=0.25$ they start to spread again. 
Second, with increasing $x_{\rm cyc}$ the sequence of the curves in each panel
changes from $0.75, 1, 1.25, 1.5$ to $1.5, 1.25, 1, 0.75$ from top to bottom.
Third, the spin gap $\Delta(T)$ decreases on increasing $x_{\rm cyc}$.
\begin{figure}[htpb]
  \centering
  \includegraphics[scale=0.55]{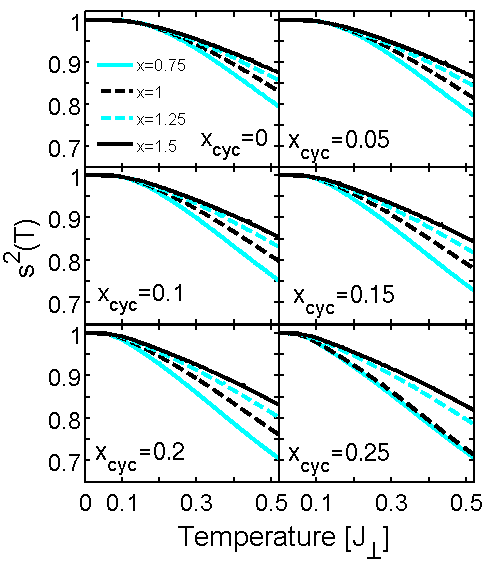}
  \caption{(color online) The dependence of the Average singlet state 
    occupation number $s^2(T)$ with temperature for various 
    $x=\left\lbrace  0.75, 1, 1.25, 1.5\right\rbrace $ with a different 
    $x_{\rm cyc}$ in each panel.}
  \label{singlis}
\end{figure}

Furthermore, it is interesting how much the average singlet state occupation 
number $s^2(T)$ changes with temperature. Its behaviour is illustrated in 
Fig.\ \ref{singlis}.  Below the characteristic temperature
$T_{\text{char}}$ the occupation number $s^2(T)$ is almost independent
of temperature  for all 
$x_{\rm cyc}$ resulting from the constant gap energy $\Delta(T)$. 
Clearly, $T_{\text{char}}$ scales with the energy gap $\Delta(0)$. Above 
$T_{\text{char}}$ the  occupation number $s^2(T)$ falls off the steeper the 
smaller $x$ for a given $x_{\rm cyc}$. 

\begin{figure}[htb!]
  \centering
  \includegraphics[scale=0.6]{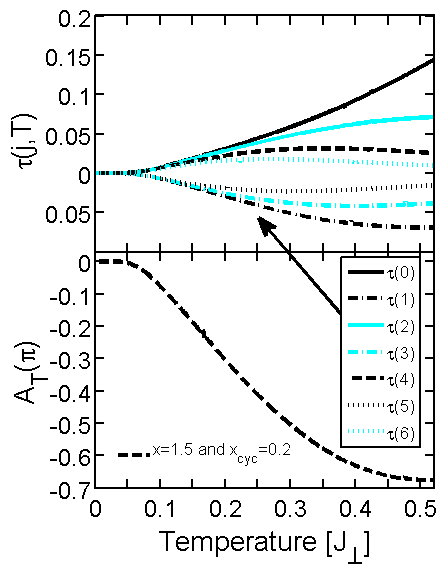}
  \caption{(color online) The upper panel shows the MF triplon hopping 
    expectation value  $\tau(j)$ for $j=\left\lbrace 0,\dots,6\right\rbrace $ 
    with $x=1.5$ and $x_{\rm cyc}=0.2$. The lower panel depicts the associated 
    conditional MF spectral weight $A_T(\pi)$.}
  \label{njApisT}
\end{figure}

The generic  behaviour of the MF triplon hopping excpectation
values $\tau(j)$ is depicted in Fig.\ \ref{njApisT} (upper panel) for 
$j=\left\lbrace 0,\cdots,6\right\rbrace$ vs.\  temperature 
for $x=1.5$ and $x_{\rm cyc}=0.2$. For all other pairs of $x$ and
 $x_{\rm cyc}$ the hopping expectation values look very similar. 
The associated conditional MF spectral weight $A_T(\pi)$ is shown in the bottom
 panel of Fig.\ \ref{njApisT}. Thus, the shape of $A_T(\pi)$ is dominated by 
$\tau(j)$ with $j\geq1$ explaining the minimum in $A_T(\pi)$ at about 
$T=0.514 J_\perp$ which is about twice
the spin gap $\Delta(0)$ \footnote{This value corresponds to about
630K for $J_\perp=105.5$meV in order to provide a first quantitative link
to experiment. This temperature corresponds to about twice the spin
gap found in Ref.\ \onlinecite{notbo07}.}.

\begin{figure}
  \centering
  \includegraphics[scale=0.45]{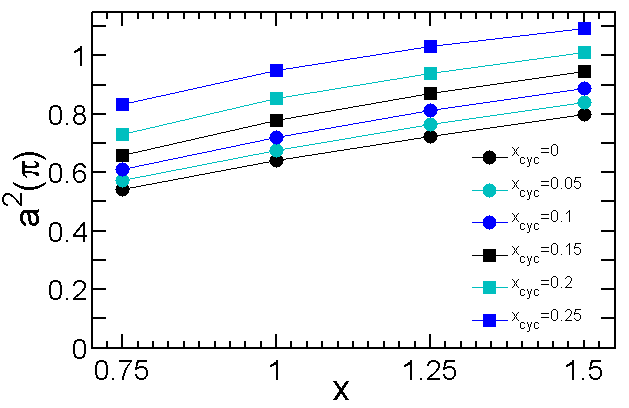}
  \caption{(color online) One-triplon spectral weight at $k=\pi$, where 
    the spin gap $\Delta(0)$ occurs, 
    for  $T=0$ vs.\ $x$ for various $x_{cyx}$.
    The  one-triplon spectral weight at $k=\pi$  increases on increasing $x$ 
    for fixed  $x_{\text{\rm cyc}}$. For fixed  $x$ the one-triplon 
    spectral weight  increases also on increasing $x_{\text{\rm cyc}}$.}
  \label{apix}
\end{figure}

\begin{figure}[htpb]
  \centering
  \includegraphics[scale=0.5]{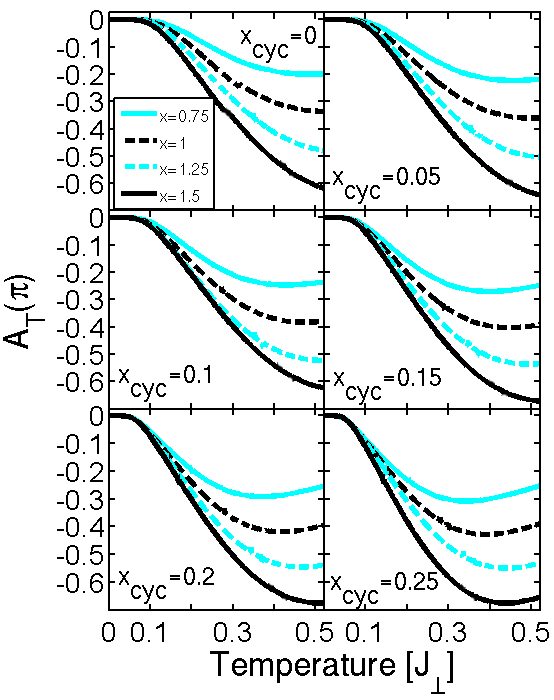}
  \caption{(color online) Conditional MF spectral weight vs.\  temperature $T$ 
    for $x=\left\lbrace 0.75, 1, 1.25,1.5\right\rbrace $ and different
    $x_\text{cyc}$ in each panel.  }
  \label{spectis}
\end{figure}

The zero temperature one-triplon spectral weight $a^2(\pi)$ vs.\ $x$ for 
$x_{\rm cyc}=\left\lbrace  0,0.05, 0.1, 0.15, 0.2, 025\right\rbrace $ 
is shown in Fig.\ \ref{apix}. With increasing $x$ the one-triplon spectral 
weight  at $k=\pi$ grows. This increasing weight on increasing $x$ 
confirms the conclusion, that the most important contribution to the one 
triplon weight and to the conditional triplon excitation weight is found at 
$k=\pi$. This is the dominant feature in the dynamic structure factor.

\begin{figure}[htb!]
  \centering
  \includegraphics[scale=0.54]{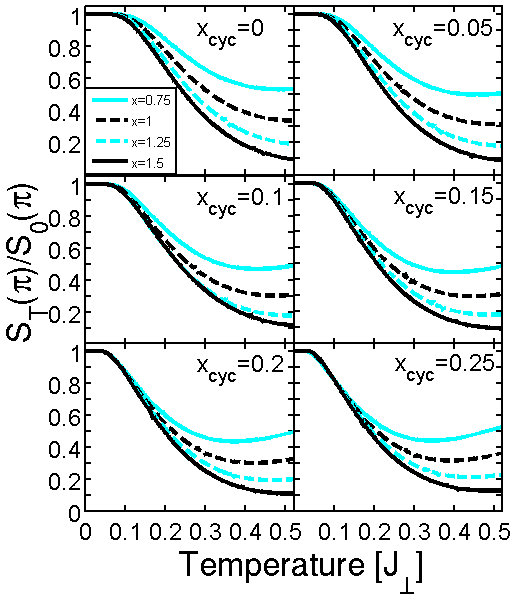}
  \caption{(color online) The momentum integrated dynamic structure factor 
    $I(T)=\int S(k, T) \text{d}k$ for $x=\left\lbrace 0.75, 1, 1.2, 
    1.5\right\rbrace$ with a different 
    $x_{\text{\rm cyc}}$ in each panel is presented. It is plotted vs.\
    temperature, normalized to the  momentum 
    integrated dynamic structure factor $I_0=\int S(k, 0) \text{d}k$ at zero 
    temperature.}
  \label{compareJperp}
\end{figure}

Fig.\ \ref{spectis} depicts the conditional MF triplon spectral weight 
$A_T(\pi)$ obtained for various  
$x=\left\lbrace 0.75, 1, 1.25,1.5\right\rbrace $ and different values
$x_\text{cyc}$ vs.\  temperature. 
The conditional MF triplon weight for various $x$ shows the expected almost 
constant behaviour below the characteristic temperature $T_{\text{char}}$ 
where all curves lie indistinguishably on top of one another. 
Above $T_{\text{char}}$ a steep fall-off is found. 
The steepness is affected by the values of $x$ and $x_{\rm cyc}$.  
The fall-off is the larger the larger $x$ and $x_{\rm cyc}$ are.
For  fixed $x_{\rm cyc}$  the fall-off becomes stronger on increasing $x$.

The calculated conditional MF amplitude $A_T(\pi)$ displays a minimum 
at about $J_\perp/2$ which corresponds to about twice the spin
gap $\Delta(0)$. We do not see any physical reason
why the weight should increase again. Furthermore, the 
justification for the approximations involved is less sound
beyond temperatures of about $2\Delta(0)$. Life time effects
of the modes are expected to become more and more important
\cite{james08,essle08,james09} so that we do not expect
the calculated data to be reliable for temperatures beyond $J_\perp/2$.

The negative conditional MF amplitude $A_T(\pi)$ diminishes the one-triplon 
spectral weight $a_T(k)^2$ according to Eq.\ (\ref{transition}) yielding the 
normalized momentum-integrated structure factor $\frac{I(T)}{I_0}$ presented 
in Fig.\ \ref{compareJperp}. This is the experimentally relevant
quantity. The shape of the normalized momentum-integrated structure factor 
$\frac{I(T)}{I_0}$ in Fig.\ \ref{compareJperp}
is very similar to the one of the conditional MF spectral weight
in Fig.\ \ref{spectis} confirming that the conditional excitation
process is the dominating effect on increasing temperature.
This fact demonstrates the importance of vertex corrections
in strongly correlated systems.

\section{Comparison to Experiment}
\label{CE}

In an inelastic neutron scattering experiment the dynamic structure factor is
 measured by the intensity of the scattered neutrons. In order to compare with 
theory a series of measurements 
were made on a cuprate spin ladder at different temperatures $T$. These 
measurements reveal the temperature development of the structure factor. 

\begin{figure}[htpb!]
  \centering
  \includegraphics[scale=0.33]{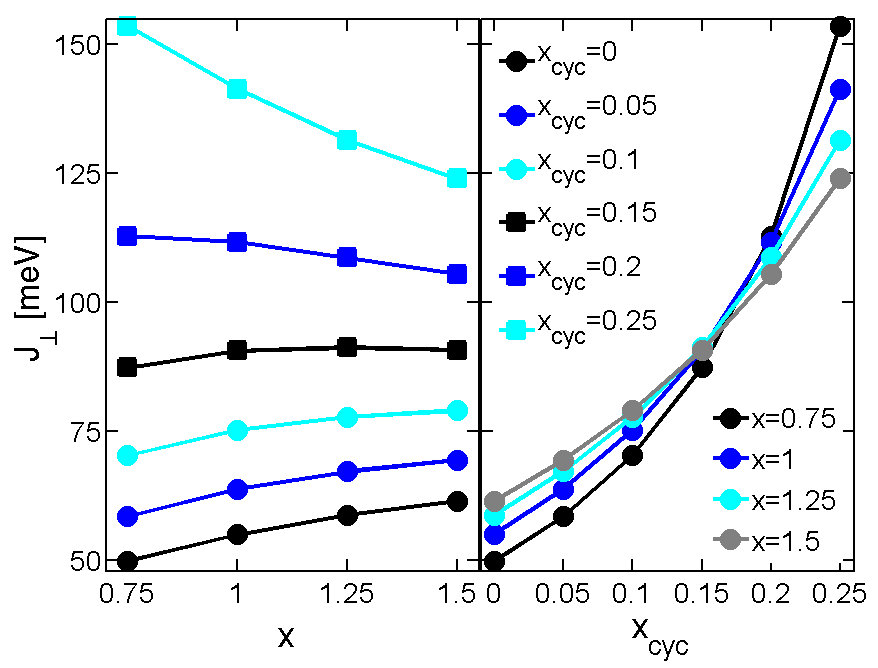}
  \caption{(color online) Rung coupling constants $J_\perp$ 
    vs.\ $x$ (left panel) and vs.\ $x_{\rm cyc}$ (right panel)
    determined from the experimental energy gap $\Delta(0)=27.6$meV 
   \cite{notiz2,notbo07}.}
  \label{couplings}
\end{figure}

So far we discussed the theoretical results relative to
the energy scale given by the  coupling constant $J_\perp$.
In order to compare to experimental data, we have to determine
this energy scale.
For each pair $\left\lbrace x, x_{\rm cyc}\right\rbrace $ the  coupling 
constant $J_\perp$ is 
determined from the experimental spin gap $\Delta(T=0)=27.6$meV
 \cite{notiz2,notbo07}. 
The obtained values of $J_\perp$ are shown in Fig.\ \ref{couplings} 
as functions of $x$ (left panel) and of $x_{\rm cyc}$ (right panel).

A striking point is found in the right panel of Fig.\ \ref{couplings}
where for $x_{\rm cyc}=0.15$ the value of $J_\perp$ is nearly 
the same for all $x$. The sequence of the curves for various values
of $x$ changes at $x_{\rm cyc}=0.15$, i.e., for lower $x_{\rm cyc}$
the curves rise on rising $x$ while they fall for larger $x_{\rm cyc}$.
Thus the curve  for $x_{\rm cyc}=0.15$ in the left panel of 
Fig.\ \ref{couplings} is almost flat.

\subsection{Sample and INS Experiments}

Single crystals of La$_{4}$Sr$_{10}$Cu$_{24}$O$_{41}$ were grown using the 
`traveling solvent floating zone' method \cite{revco99} at $9$ bar oxygen 
pressure. INS measurements were performed on the undoped ladder, 
 La$_{4}$Sr$_{10}$Cu$_{24}$O$_{41}$ 
using the MAPS time-of-flight spectrometer at ISIS, Rutherford Appleton 
Laboratory, U.K. The sample consisted of three co-aligned single crystals 
with a total mass of 23g (see  Ref.\ \onlinecite{notbo07} for details).
The crystals were mounted in a closed cycle cryostat with the $(0kl)$ 
reciprocal lattice plane horizontal and the $c$ axis perpendicular to the 
incident neutron beam $k_i$. A Fermi chopper was used to select an incident 
neutron energy of $100$meV and $50$meV and was rotated at a speed of $300$ Hz 
to give an energy resolution at the elastic line of  
$4.1$meV and $1.6$meV, respectively. Close to the gap, 
at an energy transfer of $30$meV the 
energy resolution is $3.06$meV and $0.93$meV at an incident 
neutron energy of $100$meV and $50$meV, respectively.
Data were collected 
at temperatures of $15$K, $50$K, $100$K, and $150$K, and of 
$15$K, $50$K, and $150$K  with an incident energy of $100$meV and $50$meV, 
respectively. Incoherent nuclear scattering 
from a vanadium standard was used to normalize the magnetic cross-section.
In the following text wave vectors $Q_c$ and $Q_a$ represent the direction 
along the ladder and along the rung respectively.

\subsection{Background Subtraction and Data Analysis}

\begin{figure}
  \centering
  \includegraphics[scale=0.33]{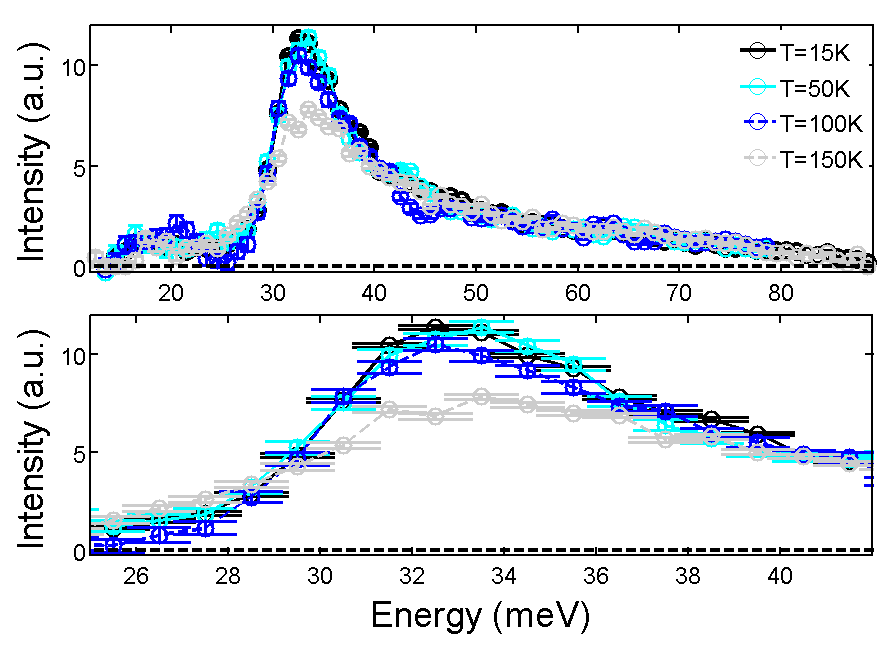}
  \caption{(color online) Background subtracted ladder signals measured at 
    various temperatures extracted from INS measurements on 
    La$_{4}$Sr$_{10}$Cu$_{24}$O$_{41}$. The lower 
    panel shows a zoom of the top panel. }
  \label{data}
\end{figure}

The ladder signal was extracted by taking a constant-$Q$ cut at $Q_c=0.5$ and 
$Q_a=1.2$ (wave vector ranges $0.45<Q_c<0.55$ and 
$0.9<Q_a<1.5$) where $Q_c$ and $Q_a$ are given in units of 
$2\pi/c$ and $2\pi/a$, respectively. 
The wave vector $Q_c=0.5$ corresponds to the position of 
the gap and the $Q_a$ range is chosen to 
maximise the ratio of magnetic intensity to background due to the rung 
modulation which goes as $\approx (1-\cos(Q_a d_\text{rung}))$, where 
$d_\text{rung}$ is the rung distance. The background was determined from a
constant-$Q$ cut for the wave 
vector range $-0.2<Q_c<0.2 $, and $0.9<Q_a<1.5 $. 
This range is appropriate for the determination of the
background, because it contains no magnetic scattering due to the ladder 
structure factor. Each temperature run was treated in exactly the same way, 
yielding the data shown  in Fig.\ \ref{data}. 

The highest intensity was measured for the coldest run at $15$K. The 
experiment does not reveal any temperature dependence between $15$K and $50$K. 
A slight decrease in intensity is observed for $100$K and a significant 
decrease for $150$K.
The spin excitation gap is found to be at $E_\Delta=(30 \pm 0.5)$meV, 
see lower panel of Fig.\ \ref{data}, 
which lies a little lower than that found in Refs.\ 
\onlinecite{eccle96,azuma94} and a little higher than found in Ref.\ 
\onlinecite{notbo07}. Here we define the gap as the 
energy below the intensity maximum where the 
scattering strength is half of its maximum value.

The FWHM of the observed peak was extracted
by fitting a gaussian to the data. As the real shape of the measured curve
deviates from the ideal gaussian, the fit is adjusted such that the
gaussian and the data points with an energy below the energy of the
maximum intensity match each another.
The FWHM values are found to be $5.3\pm0.5$meV for
$15$K, $5.1\pm0.5$meV for $50$K, $5.3\pm0.5$meV for $100$K and $(8
\pm0.5)$meV for
$150$K for an incident energy of $100$meV and $3.3 \pm0.5$meV for $15$K,
$3.5\pm0.5$meV for $50$K and $(5.8\pm0.5)$meV for $150$K for an incident
energy of $50$meV. Different FWHM values obtained for data collected at the 
same temperature but different incident energies was due to changes in the 
energy resolution.

A broadening effect due to 
damping could therefore only be observed at $150$K. This result confirms 
that the broadening with increasing temperature is small in comparison to the 
intensity decrease with increasing temperature, which in turn validates the 
MF approach presented here. Strikingly the experimental intensity curves in 
Fig.\ \ref{data} lie exactly on top of each other above $40$meV for all 
measured temperatures, indicating that there is a temperature independent 
tail which we attribute to the wave vector resolution.

\begin{figure}[htb!]
  \centering
  \includegraphics[scale=0.65]{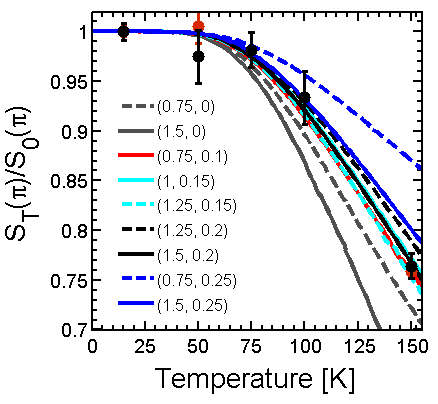}
  \caption{(color online) 
    Temperature dependence of normalized dynamic structure factor 
    $\frac{S_T(\pi)}{S_0(\pi)}$ at $k=\pi$ for various pairs of 
    $\left( x, x_{\text{\rm cyc}} \right)$. The energy scale
    is set by $J_\perp$ given in Tab.\ \ref{table}. 
    The black and red (gray) dots with error bars depict the 
    experimental result 
    obtained from INS for $100$meV and $50$meV, respectively. }
  \label{compare}
\end{figure}

The data is compared to the temperature dependent 
theoretical spectral weight described 
previously. To determine the relative changes we exclusively used the 
intensity signal in the range $28 \leq \omega \leq 40$meV and calculated the 
integrated intensities normalized by the area obtained from the $15$K run. 
Fig.\ \ref{compare} shows the comparison between 
theory and experiment for the temperature dependent changes of the normalized 
integrated intensities. The temperature scale is restricted to the temperature 
region that has been probed experimentally.

Possible pairs of $\left( x,x_{\rm cyc}\right) $ matching the INS data are 
$\left\lbrace \left( 0.75,0.1\right), \left( 1,0.15\right),
\left( 1.25, 0.15\right),\left( 1.5,0.2\right)\right\rbrace $. However, 
previous INS measurments \cite{notbo07} and optical conductivity measurements 
\cite{nunne02,schmi05b} showed, that values of $x\approx1.25-1.5$ and 
$x_{\rm cyc}\approx 0.20-0.27$ are realistic.
To be consistent with previous experimental results we argue that the best 
agreement within the experimental error bars is found for $x=1.5$ and 
$x_{\rm cyc}=0.2$ (black solid curve).

\begin{table}[h]
  \centering
  \begin{tabular}{| c | c | }
    \hline
    $\left(x,x_\text{cyc} \right)$  & $J_\perp \!\left[ \text{meV}\right] $ 
    \\ \hline
    $\left(0.75, 0 \right)$ & $54.2$ \\ \hline
    $\left(1.5, 0 \right)$ & $66.8$ \\ \hline
    $\left(0.75, 0.1 \right)$ & $76.4$ \\ \hline
    $\left(1, 0.15 \right)$ & $98.4$ \\ \hline
    $\left(1.25, 0.15 \right)$ & $99.1$ \\ \hline
    $\left(1.25, 0.2 \right)$ & $118$ \\ \hline
    $\left(1.5, 0.2 \right)$ & $115$ \\ \hline
    $\left(0.75, 0.25 \right)$ & $166.8$ \\ \hline
    $\left(1.5, 0.25 \right)$ & $135$ \\ \hline
  \end{tabular}
  \caption{Values of the coupling constant $J_\perp$ 
    based on a spin gap of $\Delta =30$meV for the 
    various pairs of  $\left(x,x_\text{cyc} \right)$ for which data
    is presented in Fig.\ \ref{compare} }
  \label{table}
\end{table}

In Ref.\ \onlinecite{notbo07}  $x=1.5$ and $x_{\rm cyc}=0.25$ with 
$J_\perp=124$meV were fitted 
to the one triplon dispersion extracted from experiment. This little 
discrepancy in $x_{\rm cyc}$ does not really matter because of the experimental
errors and the approximation necessary for computing  $A_T(\pi)$.

\begin{figure}
  \centering
  \includegraphics[scale=0.65]{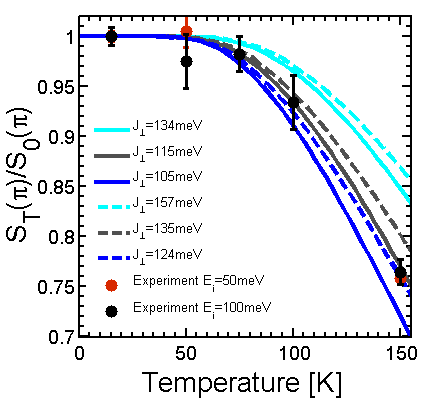}
  \caption{(color online) 
    Normalized spectral weight based on different spin gaps 
    ($\Delta= 35$meV (cyan (light gray) lines), $\Delta= 30$meV (dark lines), 
    and $\Delta= 27.6$meV (blue (dark gray) lines)) is shown. Dashed lines 
    rely on $\left(x,x_\text{cyc}\right)= \left( 1.5, 0.25\right)$ and 
    solid lines on $\left(x,x_\text{cyc}\right)= \left( 1.5, 0.2\right)$. 
    The black and red (dark gray) dots with error bars depict the 
    experimental result 
    obtained from INS for $100$meV and $50$meV, respectively.}
\label{comp}
\end{figure}

Furthermore, if slightly different energy gaps are assumed,
 such as $\Delta= 35$meV (found in Refs.\ \onlinecite{eccle96,azuma94}), 
$\Delta= 30$meV (present work) and $\Delta= 27.6$meV \cite{notiz2,notbo07}, 
slightly
different results are implied as presented in Fig.\ \ref{comp}.  Both a gap of 
$30$meV and of  $27.6$meV 
match the data for realistic parameters  $\left\{x,x_\text{cyc}\right\}$. In 
fact, using the spin gap $27.6$meV consistent with the data 
from Ref.\ \onlinecite{notbo07} 
leads to convincing agreement with the values for 
$\left\{x,x_\text{cyc}\right\}$ found in Ref.\ \onlinecite{notbo07}
to fit the INS data. Hence our findings in this work are fully
consistent with previous results.

At about half the spin gap energy $\Delta(0)$
our MF theory of vertex corrections predicts a decrease 
of the spectral weight by $25\%$ for 
$x=1.5$ and $x_{\rm cyc}=0.2$ with $J_\perp=115$meV
which  agrees very nicely with experiment, see Fig.\ \ref{comp}.
 We stress that a decrease of $25\%$ is well in the regime where
we consider our approach valid and reliable.

Note, however, that we omitted the line broadening
due to thermal fluctuations in the present analysis
which play a more and more important role on increasing 
temperature. In particular, the unique experimental determination
of the weight becomes increasingly difficult.


\section{Conclusions}

\label{Conclusions} 

In summary, we derived in this paper vertex corrections
in the dynamic structure factor for a quantum antiferromagnet
without long range order. The model system investigated is the two-leg
spin ladder. The vertex corrections are linked to thermal fluctuations;
they are relevant only at finite temperatures. We computed them
in the framework of continuous unitary transformations which lead
to an effective description of the system in terms of hardcore
triplons as elementary excitations.

The vertex corrections are evaluated on the mean-field level,
i.e., on the level of a single-model approximation. We found that 
they induce conditional excitation process which \emph{reduce} the 
unconditional, zero-temperature excitation amplitude.
Thereby, the spectral weight of the low-lying excitations
is diminished because its temperature dependence 
is indeed found to be dominated by the vertex corrections.
The vertex corrections stem eventually from the hardcore 
character of the triplons. Yet we emphasize that they go beyond the
obvious mechanism that no triplon can be excited if its
site is already excited by another triplon.

We compared the calculated spectral weight quantitatively with the
one measured in the undoped ladder La$_{4}$Sr$_{10}$Cu$_{24}$O$_{41}$ 
using inelastic neutron scattering for a range of temperatures 
up to $150$K, equivalent to  half the spin gap energy. 
The theory should still apply to considerably higher  temperatures 
as long as these stay below twice the spin gap energy.
Experiments at higher temperatures for undoped ladders could 
test this framework further.

Further research, both theoretical and experimental,
 on the vertex corrections in other
 low-dimensional quantum antiferromagnets is called for.
In particular, a challenging project consists in the 
combination of approaches studying the combined effect of line broadening
of the single modes \cite{james08,essle08,james09}
and of the vertex corrections.

\begin{acknowledgments}
We thank T.G.\ Perring, M.\ Reehuis and C.\ Gibson for experimental support. 
K.P.S.\ acknowledges ESF and EuroHorcs for funding through his EURYI. I.E.\
acknowledges the financial support by the NRW Forschungsschule 
\emph{Forschung mit Synchrotonstrahlung in den Nano- und Biowissenschaften}.
\end{acknowledgments}


\begin{thebibliography}{57}
\bibitem[{\citenamefont{note2}()}]{notiz2}
The value $\Delta= 27.6$meV results from a theoretical 
re-analysis of the 
data from Ref.\ \onlinecite{notbo07}; it agrees with the published value of 
$26.4$meV within the experimental resolution.
\end{thebibliography}


\end{document}